\documentclass[12pt]{article}

\usepackage{graphicx}

\title{ Chiral dynamics with confinement versus  the standard chiral theory }
\author{  Yu.A.Simonov \\
  NRC ``Kurchatov Institute'' -- ITEP
 \\
Moscow, 117218 Russia}

\newcommand{\beq}{\begin{eqnarray}}
 \newcommand{\eeq}{\end{eqnarray}}
\newcommand{\be}{\begin{equation}}
 \newcommand{\ee}{\end{equation}}

 \def\la{\mathrel{\mathpalette\fun <}}

\def\fun#1#2{\lower3.6pt\vbox{\baselineskip0pt\lineskip.9pt
\ialign{$\mathsurround=0pt#1\hfil ##\hfil$\crcr#2\crcr\sim\crcr}}}

\newcommand{{\SD}}{\rm SD}

\newcommand{{\Mc}}{\mathcal{M}}

\newcommand{\vex}{\mbox{\boldmath${\rm x}$}}

\newcommand{\lan}{\langle}
\newcommand{\ran}{\rangle}

\begin{document}
\maketitle
\begin{abstract}
Chiral dynamics is investigated using the chiral confining Lagrangian (CCL), previously derived from QCD with confinement interaction. Based on the calculations of the quark condensate, which is defined entirely by confinement in the zero quark mass limit, one can assert that chiral symmetry breaking  is predetermined by confinement. It is shown that CCL retains all basic relations of the standard chiral theory but enables one
to include quark degrees of freedom in the CCL. The expansion of the CCL provides  the GMOR relations and the masses and decay constants of all chiral mesons, including $\eta,\eta'$.
For the latter one needs to define a non-chiral component due to confinement, while the orthogonality condition defines the
wave functions and the eigenvalues. The resulting masses and decay constants of all chiral mesons are obtained in  good agreement with experimental and lattice data.

 \end{abstract}

 \section{Introduction}

The  fundamental concept of chiral symmetry breaking and chiral relations was understood 60 years ago \cite{1} and discussed in  the form of the $\sigma$ model \cite{2} and   formulated as
the chiral perturbation theory (CPTh) \cite{3}. Since then a lot of work was done in  understanding the physics of light chiral mesons and related problems, see \cite{4} for reviews. The main results of the chiral theory and its applications were  recently discussed in \cite{5} and in the extended review \cite{6}. The existing chiral theory, which we further quote as the standard chiral theory (SCTh), has important achievements as well as some drawbacks, which call for essential improvements and the extension of the SCTh, which will be the main topic of this paper.

First of all, the famous Gell--Mann--Oakes--Renner  (GMOR) relations \cite{7} provide a unique information on  the quark masses in connection to hadron masses. However, the decay coupling constants $f_i$ and the quark condensates $\Delta_i = | \lan \bar q_i q_i \ran |$, which enter in these relations, are to be inserted from other sources.

Secondly, the CPTh \cite{4,5,6} involves only chiral loops (and  not quark loops) and requires a number of arbitrary constants, especially in the orders $O(p^4, p^6,...)$.

The third important direction of the SCTh is the choice of the low energy chiral meson interaction, which however is not very accurate even at the lowest energy and should be improved, using other methods (e.g.  the dispersion method) at  somewhat higher energy.

One of the most prominent drawbacks of the SCTh is the  inability to describe chiral effects in the imposed magnetic field \cite{8,9,10,11,12,13,14,15,16,17,18}.
As it was shown in these lattice  measurements,  the quark   condensate $\Delta_i$, the decay constants $f_i$, and the pion mass all behave  in the magnetic field quite differently from the predictions, obtained in the SCTh.

As one can learn from all these examples, the main defect of SCTh is  due to the lack of the quark degrees of freedom, interacting via  confinement in the whole theory, and especially the lack of the
proper quark-chiral transitions, while higher order chiral effects are taken into account. The chiral symmetry breaking in SCTh and in particular the nonzero quark condensate occur spontaneously and GMOR relations are the consequence of it. In contrast to this approach we argue that the nonzero quark condensate $\Delta_i$ is a result of confinement with the string tension $\sigma$ and for zero quark masses one has a rough estimate where $\Delta_i$ is proportional to $\sigma^{\frac32}$. This fact was found analytically
long ago and will be discussed below and recently in \cite{18'} and is supported by
the lattice data, where one can see the similarity of the temperature dependent curves $\Delta(T)$ and $\sigma(T)$.

The confinement plays the most important role in  QCD, especially at low momenta, $q \la \sqrt{2\pi \sigma}\approx 1$~GeV, and provides the most part of hadron masses (see a recent review   \cite{19}), and in particular, it automatically generates CSB, since confinement is the scalar. As a consequence all chiral relations are fundamentally connected to confinement. It is a  wonderful circumstance, that some important chiral relations, e.g. the GMOR \cite{7}, have a form, where confinement is present only as the constants: $\Delta_i $ and $f_i$, which are both defined mostly by confinement.

There are old and new approaches, which tend to connect strong interaction (confinement) and  chiral physics. The most popular formalism of this kind was suggested by Nambu and Jona--Lasinio and is known as the Nambu--Jona--Lasino model (NJL) \cite{1} (see \cite{20} for more recent reviews), in which confinement is replaced by a nonlocal interaction. This model can be well compared with lattice data also for baryons \cite{20'}. There are  also numerous models, containing both chiral and nonchiral d.o.f., see e.g. \cite{21}, which contain additional parameters to be found from the lattice or experiment. These models are useful for the solution of problems in the given  direction, but do not have a general character. Thus one needs a more general approach, where one can calculate all parameters (hadron masses, decay constants,the quark condensates etc) within the single formalism.

In this paper we shall discuss the general approach in QCD, which may be called the chiral confining theory (CCTh) \cite{22,23,24,25,26,27,28,29,30,31}, which implements both chiral physics and confinement, and has a  merit of connecting these properties in one simple factor in the Lagrangian.  Moreover,  in some relations this factor does not enter and therefore one goes over to the SCTh, e.g. GMOR relations etc., while keeping the total  CCTh Lagrangian as it is, one has the full QCD with chiral d.o.f. in addition.
In this new formalism one can calculate all quantities, necessary for the explicit chiral dynamics, e.g. the decay constants $f_i$, the quark (chiral) condensates $\Delta_i$, the hadron masses $M_h$, and the hadron Green's functions and the wave functions, transition amplitudes etc.
As a good test of the new approach the CCTh was applied to the study of the
chiral effects in  the magnetic field. All three observables: quark condensate, pion and kaon decay constants and pion mass in magnetic field were calculated in the framework of CCTh without additional parameters
in \cite{27,28,29,30,31}
and found in good agreement with the lattice data in \cite{8,9,10,11,12,13,14,15,16,17,18}. It was found in these
analytic studies that the magnetic field is acting even stronger on the quark-antiquark d.o.f. and quark loops are even more important than the chiral  meson loops, and this fact was additionally supported by the
higher chiral order calculations in \cite{26}.
 The basic step in the CCTh approach was introduced first in \cite{32}, where the chiral effects were derived from the Dirac-type equation with confining interaction.  In \cite{22,23,24,25} the chiral propagators, the GMOR relations and explicit expressions  for $\Delta_i, f_i$ have  been derived, together with  excited chiral states.  The role of chiral d.o.f. in  heavy-light mesons was studied in \cite{33}. Finally in \cite{26} the general analysis of the  CCTh approach     was done in comparison to the CPTh. It was shown in \cite{26}, that the CCTh generalizes  CPTh, replacing the chiral loops by the corresponding quark loops, and allows to calculate all coefficients $L_i (i=1,2,...5)$ in the $p^4$ order of CPTh in good agreement with data.

Another important feature of the CCTh is that it automatically defines all transition elements and decay vertices of hadrons, containing chiral decay products, without new parameters. The important new development was done in \cite{34,35}, where the masses and decay widths of all scalar resonances of the first generation have been calculated in this CCTh without new parameters in good agreement with data, resolving in this way the old problem of the scalars. One of the main problem in this area is the exact definition for the chiral mesons of the $q$-$\bar q$-meson-meson amplitude which explain the final positions of all family of scalar mesons. In the case of PS
mesons the equivalent $q$-$\bar q$-meson amplitude is defined by the basic CCTh Lagrangian.

In this paper we shall derive the  general form of the CCTh Lagrangian with the account of the   $\eta,\eta'$ mesons,  assuming first the standard chiral identifications for these mesons, when
the masses are produced by the GMOR mechanism and finally introducing two types of the mass generation:
 1. due to the GMOR; 2. due to the standard confinement interaction (nonchiral) between quark and antiquark; 3. a combination of these two mechanisms.

The plan of the paper is as follows. In the section 2 we shall shortly review  the  derivation  and properties of CCTh, and  define the decay constants. In section 3 we obtain  and compare the meson masses and the decay constants in the reduced purely $SU(3)$ setting, without $\eta'$, in our version of the GMOR  with the standard values; we also calculate the decay constants and compare them with the lattice and experimental values, finding a reasonable agreement with data. In the section 4 the problem of both $\eta,\eta'$ masses and decay constants is discussed, assuming $\eta'$ to have a double  chiral and nonchiral nature. The discussion of the results and the outlook are given in the final section.

  \section{Chiral confining Lagrangian and the decay constants of chiral mesons}

The effective chiral confining Lagrangian  was derived in \cite{22,23,24} and has the form
  \be L_{eff} (M, \phi) =- N_c tr~\log (\hat{\partial} + \hat m + M \hat U).\label{1}\ee
Here the sign $tr$ implies summation over coordinates, the Dirac and flavor indices and
$\hat m$  is the diagonal mass matrix,  diag ($m_1, m_2, m_3)$,
$\hat U = \exp (i \hat \varphi \gamma_5)$, while  $M$ is the confining kernel, ensuring the confinement interaction between $q$ and $ \bar q$; in the local limit one can  consider $M(\vex)$  as the  linear interaction  $M(\vex) = \sigma |\vex|$, where $|\vex|=\lambda$ corresponds to the effective  distance $\lambda= |\vex-\vex'|$ between $q(\vex)$ and $\bar q(\vex')$ at the  moment of the
 $q \bar q $ - the chiral meson transition. (Note that $M U$ is always appears at the transition point). This transition distance $\lambda$ depends on the transition vertex and was calculated for the case of scalar  mesons in \cite{34,35} to be around 0.4~fm, and for the PS channel one has $\lambda= 0.2$~fm and $M(\lambda)=\sigma\lambda= 0.15$~ GeV.
\footnote{Note, that in the previous papers \cite{22,23,24} $M(\lambda)$ was denoted as $M(0) = \sigma \lambda =0.15$~GeV.}

At this point we combine all nine PS mesons, including $\eta,\eta'$ in one matrix $\hat U$.  The matrix $\hat U$  in the large $N_c$ limit, which is a  reasonably good approximation of QCD,
can be written as  in \cite{4,6},
 \be \hat U(\varphi) =\exp (i \gamma_5 \hat
\Phi),~~ \hat \Phi = \frac{\eta_1 }{f_1} \sqrt{\frac23} I_3+\hat \varphi. \label{2}\ee

Here $I_3$ is a diagonal unit matrix.

 \be
\hat \varphi = \sqrt{2}\left(
\begin{array}{ccc} \frac{\eta_8}{\sqrt{6}} +
\frac{\pi^0}{\sqrt{2}} , &\frac{\pi^+}{F},&
\frac{K^+}{F}\\
\frac{\pi^-}{F} &   \frac{\eta_8}{\sqrt{6}F}-
\frac{\pi^0}{\sqrt{2}F}, & \frac{K_0}{F}\\
\frac{K^-}{F},& \frac{\bar K_0}{F}, & - \frac{2}{\sqrt{6}}\frac{\eta_8}{
F}\end{array}\right).\label{3}\ee

Here $F$ is a temporary parameter which will be finally replaced by actual decay constants found from the
solution of the basic equations.
Our next purpose is to derive the GMOR  relations from (\ref{1}) and then the explicit  expressions for the decay constants of all PS mesons.
Following \cite{26} we shall separate out the terms, containing the quark masses $\hat m$, and write
\be L_{eff} (M,\hat \Phi) =- N_c tr \log (1-\xi),\label{4}\ee
\be \xi = \hat U ^+ \Lambda (\hat \partial+ \hat m) (\hat U - 1) = \xi_\varphi + \xi_m,\label{5}\ee
\be \xi_\varphi = \hat U^+ \Lambda\hat \partial \hat U, ~~ \xi_m = \hat U^+ \Lambda \hat m(\hat U -1).\label{6}\ee

Here $\Lambda= \frac{1}{\hat \partial +m_i + M}$.
To  the  second order in $\hat \Phi$ one obtains from (\ref{4}), expanding in $\xi$,

\be L_{eff}^{(2)} (M, \hat \Phi) =   N_c tr (\xi+ \frac12\xi^2 )\label{7}\ee

\be \xi  = i \Lambda \hat \partial \hat \Phi \gamma_5 - \Lambda m \frac{\hat \Phi^2}{2}+ \hat \Phi \bar \Lambda \hat m \hat \Phi+i\Lambda \hat m \hat \Phi \gamma_5\label{8}\ee
and finally

\be L_{eff}^{(2)} (M, \hat \Phi) =  N_c tr \left\{\Lambda \hat   m \frac{\hat \Phi^2}{2}+\frac12   \Lambda \hat\partial \hat \Phi\bar \Lambda \hat \partial  \hat \Phi -\frac12   \Lambda \hat m \hat \Phi\bar \Lambda \hat m  \hat \Phi\right\}. \label{9}\ee

Now defining the quark condensate $\Delta_i$,
\be \Delta_i \equiv |\lan \bar q_i  q_i \ran | = N_c tr_D \Lambda_i = N_C tr_D \left( \frac{1}{\hat \partial + m_i + M} \right)_{xx},\label{10}\ee
one can write the first term on the  r.h.s. of (\ref{9}) as
$$ N_c tr  \left( \Lambda \hat   m \frac{\hat \Phi^2}{2}\right) =  \frac12 \Delta_i m_i \hat \Phi_{ik} \hat \Phi_{ki} = \frac12 \left\{ \Delta_1 m_1 (\Phi^2_{11} + \Phi_{12}\Phi_{21}+ \Phi_{13} \Phi_{31})+\right.$$
$$ \left. +\Delta_2 m_2 (\Phi_{21} + \Phi_{12}+\Phi_{22}^2+ \Phi_{23} \Phi_{32})+\Delta_3 m_3 (\Phi_{31}  \Phi_{13}+\Phi_{32}  \Phi_{23}+\Phi_{33}^2) \right\}=$$
$$= \frac12 (\Delta_1 m_1+\Delta_2 m_2+\Delta_3 m_3)\left( \frac23\frac{\eta^2_1}{F^2} + \frac16 (\Delta_1 m_1 + \Delta_2 m_2 +4 \Delta_3 m_3)\frac{\eta^2_8}{F^2} \right) + \frac12 (\Delta_1 m_1+ \Delta_2m_2) \frac{(\pi^0)^2}{F^2}+$$
$$ \frac12 (\Delta_1 m_1-\Delta_2 m_2)\left( \frac{2}{\sqrt{3}}\frac{\pi^0\eta_8}{F^2} + \frac{4}{\sqrt{6}} \frac{\pi^0\eta_1}{F^2} \right) + \frac{2}{3\sqrt{2}}  \frac{\eta_8 \eta_1}{ F^2} (\Delta_1 m_1+ \Delta_2m_2- 2 m_3\Delta_3)$$
\be + \frac{\pi^+\pi^-}{F^2}   (\Delta_1 m_1+\Delta_2 m_2 ) + \frac{K^+K^-}{F^2}  (\Delta_1 m_1+ \Delta_3m_3)+  \frac{(K^0)^2}{F^2} (\Delta_2 m_2+ \Delta_3m_3).\label{11}\ee

On the other hand one can represent the quadratic meson terms in the  Lagrangian as
\be L^{(2)}_{eff} = \int
\left \{ \frac12 \sum (-\mu^2_a\tilde \varphi^2_a) + \frac12 \sum  ( (\partial_\mu \tilde  \varphi_{a})^2)  \right \} d^4 x, \label{12}\ee
where we have used the orthogonalized physical hadron states $\tilde \varphi_{\pi^0}, \tilde \eta, \tilde \eta'$ to be defined below.

At the first stage we disregard the mixing terms and comparing (\ref{11}) and (\ref{12}), obtain the GMOR relations,

\be (\tilde \pi^0)^2 \mu^2_{\pi^0} = \frac{\Delta_1 m_1 + \Delta_2 m_2}{F^2} (\pi^0)^2 \label{13}\ee

\be (\tilde \pi^+)^2 \mu^2_{\pi^+} = (\tilde \pi^-)^2 \mu^2_{\pi^-}=\frac{\Delta_1 m_1 + \Delta_2 m_2}{F^2} (\pi^+)^2 \label{14}\ee

\be (\tilde K^+)^2 \mu^2_{K^+} = (\tilde K^-)^2 \mu^2_{K^-}= \frac{\Delta_1 m_1 + \Delta_3 m_3}{F^2} (K^+)^2\label{15}\ee

\be (\tilde K^0)^2 \mu^2_{K^0} = \frac{\Delta_2 m_2 + \Delta_3 m_3}{F^2} (K^0)^2\label{16}\ee

\be (\tilde \eta)^2 \mu^2_\eta = \frac{(\Delta_1 m_1 + \Delta_2 m_2+ 4 \Delta_3 m_3)}{3F^2} (\eta_8)^2 \label{17}\ee

\be (\tilde \eta)^2\mu^2_{\eta_1} = \frac{2 (\Delta_1 m_1 + \Delta_2 m_2+ \Delta_3 m_3)}{3F^2} (\eta_1)^2.\label{18}\ee
We are treating here these relations, without the mixing terms in (\ref{11}) to simplify the discussion and we shall take them into account in the final equations.
As it is, we do not yet have the actual GMOR relations, since we have one common decay constant $F$ and the ratio of the normalization of the initial $(\phi_a)$ and physical $ (\tilde \phi_a) $ wave functions is not yet defined. This ratio can be obtained from the definition of the real decay constants, which should replace the parameter $F$.

We now turn to the fundamental problem of computing the meson decay constants $f_i$, following \cite{26,36,37,38}. To this end we consider the second term on the  r.h.s. of (\ref{9}), which should have  the standard form as in (\ref{12}). Writing $\Lambda_i = \bar \Lambda_i^{-1} G_i,~~ G_i=\Lambda_i \bar \Lambda_i,$, one obtains
\be \frac{N_c}{2} tr_{Dfx} (\Lambda \hat \partial \hat \phi \hat \Lambda \hat \partial \hat \phi) = \frac{N_c}{2} tr_{Dfx}\left\{\bar \Lambda_i^{-1} G_i \hat \partial \Phi_{ik} \Lambda_k^{-1} G_k\hat \partial \Phi_{ki}\right\}.\label{19}\ee

Finally, taking into account the equations of motion,  one obtains, as in \cite{36,37,38},
\be f^2_{ik} = N_c(m_i +M(\lambda)) (m_k + M(\lambda)) \int \frac{d^4p}{(2\pi)^4} G_i (p) G_k(p) = \frac{ N_c (m_k + M(\lambda)) (m_i + M(\lambda))}{\omega_i \omega_k {\cal{M}}_{ik} \xi_n} |\psi_{ik}(0)|^2,\label{20}\ee
where $\omega_i $ is the average energy of the quark $i$  and $\psi_{ik}$ is the quark-antiquark wave function at the origin (see \cite{36,37,38} for  explicit calculations and definitions).

The next step is  the  attributing $f_{ik}$ to the physical hadron decay constants $f_\pi,  f_K$ etc.  To this end we are writing in (\ref{19})
$$\frac14 \int \partial_\mu \Phi_{ik}(x) \partial_\mu \Phi_{ki}(x) f^2_{ik} d^4x = \frac{1}{2F^2} \partial_\mu \pi^+ \partial_\mu \pi^-  f^2_{12}  + \frac{1}{2F^2} \partial_\mu  K^+ \partial_\mu  K^-  f^2_{13} +$$
$$ + \frac{f^2_{23}}{2F^2} (\partial_\mu  K^0)^2 + \frac{1}{12} (f^2_{11} + f^2_{22} + 4 f^2_{33}) \frac{(\partial_\mu \eta_8)^2}{F^2}+  $$
\be \frac{1}{2\cdot 3} (f^2_{11} + f^2_{22} +  f^2_{33}) \frac{(\partial_\mu \eta_1)^2}{F^2}+ \frac14 \frac{(\partial_\mu  \pi^0)^2}{F^2} (f^2_{11} + f^2_{22} ).\label{21}\ee

As a result one can connect the physical meson wave functions $\tilde \phi_a$ with the initial wave functions, listed in the $SU(3)$ complex in (\ref{3}), while $\phi_a$ (the ninth PS meson $\eta'$)  requires additional analysis, done in the next section. Comparing (\ref{12}) and (\ref{21}), one finds
\be
\tilde \phi_a(x)= \phi_a(x) \frac{f_{a}}{F}. \label{22}\ee

Here $\phi_a(x)$ are the initial chiral mesons $\pi(x), K(x), ...$.
In addition one can define the decay constants of the physical mesons, which enter the GMOR relations and are internally connected with the corresponding physical wave functions, i.e. the connection  of $f_{ik}$ and $f_\pi, f_k$... Namely,
 \be
 f^2_{\pi^+} = f^2_{12}, ~ f_{K^+} = f^2_{13}, ~~ f^2_{K^0}= f^2_{23}, ~~ f^2_{\pi^0} = \frac12   (f^2_{11} + f^2_{22} ) \label{23}\ee
\be f^2_{\eta'}  = \frac13  (f^2_{11} + f^2_{22}+ f^2_{33}), ~~ f^2_\eta = \frac16 (f^2_{11}+ f^2_{22}+ 4 f^2_{33}).\label{24}\ee

Finally one should take into account the last term on the r.h.s. of  (\ref{9}), which contributes $O(m^2)$ corrections to the hadron masses and can be written as in \cite{26},
\be \mu_a^2 = \bar \mu^2_a - (\Delta\mu_a)^2 , ~~ a= \pi, K,...\label{25}\ee
and $(\Delta \mu_a)^2$ are given in \cite{26}, here we only add
\be (\Delta \mu_{\eta_1})^2 = 2 (\Delta \mu_\eta)^2 = \frac13 (m^2_1+m^2_2+ 4m^2_3).\label{26}\ee

\section{The masses and decay constants of the chiral mesons in the pure SU(3) approximation}

In this section we neglect any connection with the $\eta'$ meson and consider only eight PS mesons in the
SU(3) matrix in (\ref{3}). To define their masses and decay constants from the Eqs.~(\ref{13})--(\ref{18})
with account of the relations (\ref{21})--(\ref{22}) one needs three sets of the values: $ m_1, m_2, m_3 ; \Delta_1, \Delta_2, \Delta_3; f_{11}, f_{12}, f_{22}, f_{23}, f_{33}$. For the quark masses one can use the values, which are in the corresponding accepted intervals \cite{39} e.g.: $ m_1= 5, m_2= 9, m_3= 150 $ (all in MeV) at the scale  $1$~GeV.
As it was discussed in the Introduction, the quark condensates $\Delta_i$ are nonzero due to confinement and
can be calculated via the string tension $\sigma$, as in \cite{18',24}. In the following we shall also use the lattice data, e.g. from \cite{39},
\be
 \Delta_1= \Delta_2= \Delta_l= (283~{\rm MeV})^3, \Delta_3= (290~{\rm MeV})^3.
 \label{27} \ee
 Now using (\ref{13}) and (\ref{22}), one obtains the GMOR relations, containing only physical dynamical variables and the constants:
 \be
 f_l^2 \mu_{\pi^0}^2= m_1 \Delta_1 + m_2 \Delta_2, f_K^2 \mu_K^2= \Delta_1 m_1 + \Delta_3 m_3.
 \label{28} \ee

Finally the decay constants $f_{ik}$ can be calculated analytically, using (\ref{20}), as shown in \cite{37,38,39}. Then with the 10 percent accuracy one has  $f_{11}= f_{22}= f_{12}= f_l= 133$ MeV, in good agreement with the experimental $f_\pi^{ex}=131$ MeV, which provides the pion mass with the accuracy around $10\%$.

Similarly for the $K$ meson from (\ref{20}) one obtains the ratio $\frac{f_K}{f_\pi}= 1.209$, which agrees well
with the experimental values $\frac{f_K}{f_\pi}= 1.22\pm 0.02 $ \cite{41}.

We now turn to the eighth meson $\eta_8$, which in our SU(3) approach is identified with $\eta$,  and using (\ref{17}) and (\ref{24}) one can write
\be
         f_\eta^2= \frac16 (f_{11}^2 + f_{22}^2 + 4 f_{33}^2) \approx f_\pi^2, f_\eta^2 \mu_\eta^2=
 \frac13 (m_1 \Delta_1 + m_2 \Delta_2 + 4 m_3 \Delta_3).
\label{29}\ee

Now one can estimate the mass of the $\eta$ with the accuracy of 10 percent (neglecting $m_1,m_2$ as
compared to $m_3)$
\be
m_{\eta}^2= \frac{m_1 \Delta_1 + m_2 \Delta_2 + 4 m_3 \Delta_3}{3 f_{\eta}^2}= 0.29 ~{\rm GeV}^2, m_{\eta}= 0.54 ~{\rm GeV}.
\label{30}\ee

This result for $\eta$ is close to the experimental value, while  $\eta'$ was assumed above to be of nonchiral formalism  -- without a possible mixing with $\eta $. Now we turn to all nine mesons including
$\eta'$ and will assume a possible mixing with $\eta$.

\section{The $\eta-\eta'$ mixing and their resulting masses}

At this point the problem of $\eta'$ appears. The purely nonchiral Hamiltonian for the PS meson with account of $s\bar s$ it yields mass
of $\eta'$ in the range $750-770$ MeV \cite{40}, which is far from the experimental value.  One way to solve this problem is to assume that $\eta'$ belongs to the same chiral set of states as a chiral scalar, where it gets its mass  from the (chiral) GMOR relations, as it is assumed in the CPTh \cite{40}, but also obtains ite mass from confinement as all nonchiral mesons. In this case, using (\ref{11}) and neglecting tiny mixing with $\pi^0$, one obtains two equations
\be
\frac{m_1 \Delta_1 + m_2 \Delta_2}{2 F^2} \left(\eta_1 \sqrt{\frac{2}{3}} + \eta_8 \frac{1}{\sqrt{3}}\right)^2 + \frac{\Delta_3 m_3}{2F^2}\left(\eta_1 \sqrt{\frac{2}{3}}- \eta_8\frac{2}{\sqrt{3}}\right)^2= \frac12 (\mu_{\eta}^2 \eta^2 + \mu_{\eta'}^2 \eta'^2),
\label{31}\ee
  \be
\frac{f_{11}^2 + f_{22}^2}{2F^2}
\left(\partial_\mu (\eta_1 \sqrt{\frac{2}{3}} + \eta_8 \frac{1}{\sqrt{3}})\right)^2
+ \frac{f_{33}^2}{2F^2}
 \left(\partial_\mu (\eta_1 \sqrt{\frac{2}{3}}- \eta_8 \frac{2}{\sqrt{3}})\right)^2
 = \frac12
 \left((\partial_\mu \eta)^2 + (\partial_\mu \eta')^2\right).
\label{32}\ee
One can try to solve these equations, expressing $\eta,\eta'$ as linear combinations of $\eta_8,\eta_1$
but it seems to be impossible to get rid of mixing terms and thus make these states orthogonal. At this point one can doubt the very idea that $\eta_1,\eta'$ are purely chiral objects ( getting their
masses from the GMOR) and instead one should implement the idea of a pure $q \bar q$ meson like $\phi$, also
taking into account that both have similar masses.

To this end we are adding the standard confinement generated mass of the $s\bar s$ PS state $M_1$ to the (\ref{31}) and obtain the new equation
\be
\frac12 M_1^2 \eta_1^2 + \frac{m_1 \Delta_1 + m_2 \Delta_2}{2 F^2} \left(\eta_1 \sqrt{\frac{2}{3}} + \eta_8 \frac{1}{\sqrt{3}}\right)^2 + \frac{\Delta_3 m_3}{2F^2}\left(\eta_1 \sqrt{\frac{2}{3}}- \eta_8 \frac{2}{\sqrt{3}}\right)^2= \frac12 (\mu_{\eta}^2 \eta^2 + \mu_{\eta'}^2 \eta'^2),
\label{33}\ee
while the second equation (\ref{32}) is kept as it is. Now one must use these equations with addition of (\ref{22}) to define the orthogonalized combinations of $\eta_1,\eta_8$ for the final $\eta,\eta'$. We write $\eta_1= \sin\psi_1 \eta + \cos\psi_1 \eta', \eta_8= \cos\psi_2 \eta + \sin\psi_2 \eta'$.
Now from the (\ref{32}), (\ref{33} ) one defines the coefficients of the mixed terms in both equations and from the (\ref{32}) one obtains $\psi_1= - \psi_2$, while from
(\ref{33}) one finds
\be
tg2\psi_1= 2 \sqrt2 \frac{2\Delta_3 m_3}{3 F^2 M_1^2}.
\label{34}\ee
In (\ref{34}) we have neglected the contribution of $(\Delta_1 m_1 + \Delta_2 m_2)$ as compared to $\Delta_3 m_3$, which is around 1/10,  and replacing $F$ by $f_{\pi}$  our results will have the accuracy around 10 percent. Now one must insert the nonchiral singlet $s\bar s$ mass, which is the PS equivalent of the $\phi$ meson with the mass $1.02$ GeV. Using the relativistic formalism with confinement, which allows to predict the majority of light, heavy, and heavy-light  \cite{40}, one can predict the mass of the PS meson
in the region $750-780$ MeV. Inserting this value in (\ref{34}) one obtains $\sin\psi_1= 0.286$ and the expressions for $\mu_{\eta},\mu_{\eta'}$ acquire the form
$$
\mu_{\eta}^2= M_1^2 (\sin\psi_1)^2 + \frac{\Delta_3 m_3}{F^2} \left(\frac23 \sin^2\psi_1 + \frac43 \cos^2\psi_1 + \frac{2\sqrt2}{3} \sin^2\psi_1 \right)
\mu_{\eta'}^2=$$\be= M_1^2 (\cos\psi_1)^2 + \frac{\Delta_3 m_3}{F^2} \left(\frac23 (\cos\psi_1)^2 + \frac43 (\sin\psi_1)^2 +
\frac{2\sqrt2}{3} \sin2\psi_1\right).
\label{35}\ee
Inserting the values for $\psi_1$ and taking $M_1= 0.75$ GeV one obtains accuracy of around 10-15 percent
$ \mu_{\eta}= 0.44$ GeV, $\mu_{\eta'}= 0.867$ GeV, which should be compared with the experimental values
$0.55$ and $0.96$ respectively.

To obtain more realistic results we shall use for $F$ not the pion decay constant $f_{\pi}$, as above, but the realistic value of the $s\bar s$ constant $f_{ss}$, which is close to $F_0,F_8$ computed on the lattice in \cite{41} and equal to $100.1;115$ MeV. Inserting this value in (\ref{34}) one gets for the masses of $\eta,\eta'$
\be
\mu_{\eta'}= 976 ~{\rm MeV}~ ,~~ \mu_{\eta}= 514~{\rm MeV}.
\label{36}\ee
Their values are close to the experimental values cited above and to the lattice data in \cite{41}.

\section{Conclusions and the outlook}

We have considered the chiral symmetry breaking in the general QCD setting, where confinement plays the major role at low and intermediate energies. In the standard QCD approach it is believed that in this region one meets the new phenomenon -- the spontaneous chiral symmetry breaking, characterized by the nonzero value of the quark
condensate $\lan q\bar q\ran$. We have tried to look behind this phenomenon and found that the nonzero quark condensate occurs due to confinement and is roughly proportional to the string tension in the power $3/2$. This conclusion is supported by the numerous analytic and lattice calculations and is supported by the roughly simultaneous disappearance of the string tension and the quark condensate at the critical temperature. Thus the  spontaneous chiral symmetry breaking is actually predetermined by the confinement. Nevertheless all qualitative conclusions of the standard chiral symmetry breaking are kept intact, namely, the GMOR, approximate $SU(3)$ symmetry of the resulting hadron interactions etc.

However, at this point the similarity of the old CPTh and the new confining chiral symmetry breaking ends up and one discovers many places, where confining chiral symmetry breaking allows to improve the results of CPTh. The main defect of CPTh, as follows from our studies, originates from the fact that quark degrees of freedom are not properly included in CPTh and
including higher orders in the chiral Lagrangian, one is not taking into account the $q\bar q$ -- chiral degrees of freedom. This is crucial for the resulting behavior  in the magnetic field and while the CPTh predictions for the field dependence of quark condensate, decay constants and the pion mass fail \cite{8,9,10,11,12,13,14,15,16,17,18}, the new confining
chiral theory, formulated in \cite{22,23,24,25,26,31,32}, solves all these three conflicts \cite{27,28,29,30}.

In the paper we have discussed the old problem of the $\eta,\eta'$ mesons, which have been treated
in the framework of the CPTh and beyond (see \cite{42,43} for recent reviews). Here the main emphasis was done on the assumption that the GMOR type mass generation is valid for both mesons and on the search for the appropriate combination of $\eta_8,\eta_1$ units to ensure the observed masses and decay constants. Here we have also suggested  a more
general approach, where $\eta'$ is largely non-chiral meson, which acquires the chiral components due to the
orthogonality with the $\eta$ meson wave function, which uniquely defines the weights of all components, and the resulting mass values come out close to the experimental.

In this way the proposed approach can be considered as a useful extension of the standard chiral theory, which enables one to overcome old difficulties and understand the chiral theory as a necessary ingredient of the whole QCD theory including confinement.

The author is grateful to A. M. Badalian for useful discussions and  is also indebted to N. P. Igumnova for
important help in the preparing of the manuscript.

  \end{document}